# Aberration-free ultra-thin flat lenses and axicons at telecom wavelengths based on plasmonic metasurfaces

Francesco Aieta[1, 2], Patrice Genevet[1, 3], Mikhail A. Kats[1], Nanfang Yu[1], Romain Blanchard[1], Zeno Gaburro[1, 4] and Federico Capasso[1]

[1]School of Engineering and Applied Sciences, Harvard University, Cambridge, Massachusetts 02138, USA
[2]Dipartimento di Scienze e Ingegneria della Materia, dell'Ambiente ed Urbanistica, Università Politecnica delle Marche, via Brecce Bianche, 60131 Ancona, Italy
[3]Institute for Quantum Studies and Department of Physics, Texas A&M University, College Station, Texas 77843, USA
[4]Dipartimento di Fisica, Università degli Studi di Trento, via Sommarive 14,38100 Trento, Italy

**The concept of optical phase discontinuities is applied to the design and demonstration of aberration-free planar lenses and axicons, comprising a phased array of ultrathin subwavelength spaced optical antennas. The lenses and axicons consist of radial distributions of V-shaped nanoantennas that generate respectively spherical wavefronts and non-diffracting Bessel beams at telecom wavelengths. Simulations are also presented to show that our aberration-free designs are applicable to high numerical aperture lenses such as flat microscope objectives.**

The fabrication of lenses with aberration correction is challenging in particular in the mid- and near-infrared wavelength range where the choice of transparent materials is limited. Usually it requires complex optimization techniques such as aspheric shapes or multi-lens designs [1,2], which are expensive and bulky.

Focusing diffracting plates offer the possibility of designing low weight and small volume lenses. For example the Fresnel Zone Plate focuses light by diffracting from a binary mask that blocks part of the radiation [1]. A more advanced solution is represented by the Fresnel lens, which introduces a gradual phase retardation in the radial direction to focus light more efficiently. By limiting the absorption losses and gathering oblique light more efficiently, Fresnel lenses are advantageous for optical systems with high numerical aperture (NA) [1]. To guarantee a smooth spherical phase profile responsible for light focusing, the Fresnel lens thickness has to be at least equal to the effective wavelength $\lambda_{eff} = \lambda/n$ where n is the refractive index of the medium. Moreover, the thickness of the lens needs to be continuously tapered, which becomes complicated in terms of fabrication [3].

In the microwave and mm-wave regimes, local control of the phase of electromagnetic waves obtained with reflectarrays or frequency selective surfaces has enabled alternative designs for flat lenses. For example, in reflectarrays, scattering units comprising metallic patch antennas coupled with a ground plane can provide an arbitrary phase shift between the scattered light and the incident light [4, 5].

At optical frequencies, planar focusing devices have been demonstrated using arrays of nanoholes [6], optical masks [7-9], or nanoslits [10]. These techniques require complex design rules and do not provide the ability to tailor the phase of the transmitted light from 0 to $2\pi$, which is necessary for a complete control of the optical wavefront. In addition flat metamaterials based lenses such as hyperlenses and superlenses have been used to achieve subdiffraction focusing [11-14].

The concept of optical phase discontinuities, which has been used in the demonstration of new metasurfaces capable of beaming light in directions characterized by generalized laws of reflection and refraction [15], provides a different path for designing flat lenses. In this approach, the control of the wavefront no longer relies on the phase accumulated during the propagation of light, but is achieved via the phase shifts experienced by radiation as it scatters off the optically thin array of subwavelength-spaced resonators comprising the metasurface. This strategy resembles reflectarrays and transmit-arrays used at much lower frequencies [16-18]. Linear gradients of phase discontinuities lead to planar reflected and refracted wavefronts [15, 19, 20]. On the other hand, nonlinear phase gradients lead to the formation of complex wavefronts such as helicoidal ones, which characterize vortex beams [21].

In this paper, we experimentally demonstrated light focusing in free space at telecom wavelength λ=1.55 μm using 60nm thick gold metasurfaces. We fabricated two flat lenses of focal distances 3cm and 6cm and a flat axicon with an angle β=0.5° (which correspond to a glass plano-convex axicon with base angle 1°). Axicons are conical shaped lenses that can convert Gaussian beams into non-diffracting Bessel beams and can create hollow beams [22-26]. Our experiments are in excellent agreement with numerical simulations; our calculations also point to the possibility of achieving high numerical aperture lenses.

The design of our flat lenses is obtained by imposing a hyperboloidal phase profile on the metasurface. In this way, secondary waves emerging from the latter constructively interfere at the focal plane similar to the waves that emerge from conventional lenses [1]. For a given focal length $f$, the phase shift $\varphi_L$ imposed in every point $P_L(x,y)$ on the flat lens must satisfy the following equation (Fig 1a):

$$\varphi_L(x,y) = \frac{2\pi}{\lambda}\overline{P_L S_L} = \frac{2\pi}{\lambda}\left(\sqrt{(x^2+y^2)+f^2}-f\right), \quad \text{Eq.1}$$

where λ is the wavelength in free space.

For an axicon with angle β, the phase delay has to increase linearly with the distance from the center, creating a conical phase distribution. The phase shift $\varphi_A$ at every point $P_A(x,y)$ has to satisfy the following equation

$$\varphi_A(x,y) = \frac{2\pi}{\lambda}\overline{P_A S_A} = \frac{2\pi}{\lambda}\sqrt{(x^2+y^2)}\sin\beta. \quad \text{Eq.2}$$

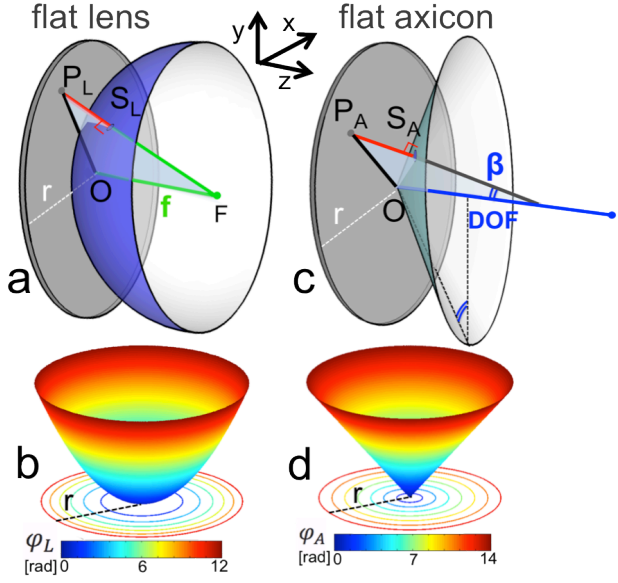

**Fig 1.** Schematic showing the design of flat lenses and axicons. In order to focus a plane wave to a single point at distance f from the metasurface, a hyperboloidal phase profile must be imparted onto the incident wavefront. (a) The phase shift at a point $P_L$ on the lens surface is proportional to the distance between $P_L$ and its corresponding point $S_L$ on the spherical surface of radius equal to the focal length f. The resulting hyperboloidal radial phase distribution on the flat lens is shown in (b); c) The axicon images a point source onto a line segment along the optical axis; the length of the segment is the depth of focus (DOF). The phase in point $P_A$ on the flat axicon is proportional to the distance between $P_A$ and its corresponding point $S_A$ on the surface of a cone with the apex at the intersection of the metasurface with the optical axis and base angle $\beta = \tan^{-1}\left(\frac{r}{DOF}\right)$, where r is the radius of the metasurface. The resulting conical radial phase distribution on the flat axicon is shown in (d). The phase profiles for the flat lenses and axicons are implemented using V-shaped optical antennas

Optical antennas with equal scattering amplitudes and phase coverage over the whole $2\pi$ range are necessary for designing flat lenses with a large range of focal distances. Following the approach previously discussed [15, 19, 21], we design 8 different plasmonic V-shaped antennas that scatter light in cross-polarization with relatively constant amplitudes and incremental phase of π/4 between neighbors. Figure 2a shows the cross-polarized scattering amplitudes and the corresponding phase shifts for the 8 elements obtained with full wave simulations using the Finite Difference Time Domain technique (FDTD). Using Eqs 1 and 2, we design two lenses with radius r=0.45mm and focal lengths f=3cm (NA = 0.015) and f=6cm (NA = 0.075), respectively, and an axicon with the same radius and an angle β=0.5°. The devices are fabricated by patterning a double-side-polished undoped silicon substrate with gold nano-antennas using electron beam lithography (EBL). The measurement setup is shown in Fig. 2b.

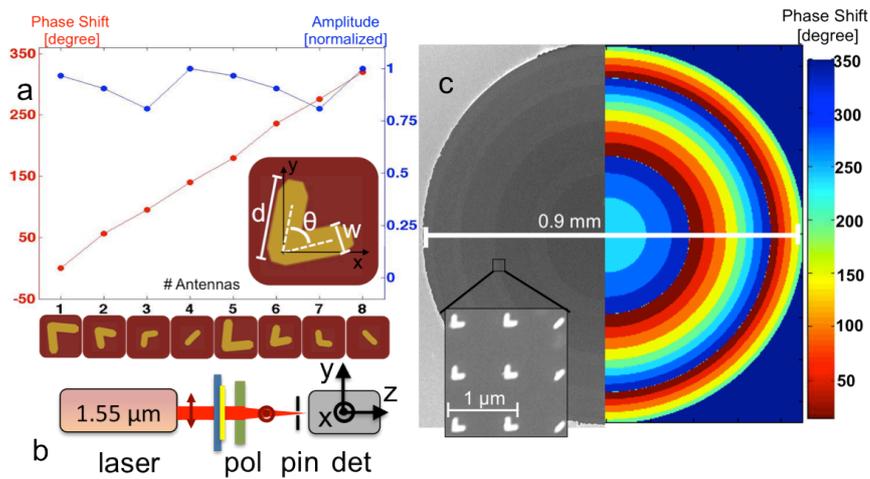

**Fig 2.** a) FDTD simulations are used to obtain the phase shifts and scattering amplitudes in cross-polarization for the 8 elements used in our metasurfaces (See Supplemental Material for details). The parameters characterizing the elements from 1 to 4 are: d = 180nm, 140nm, 130nm, 85nm and θ = 79°, 68°, 104°, 175°. Elements from 5 to 8 are obtained by rotating the first set of elements by an angle of 90° counter-clockwise. The width of each antenna is fixed at w=50nm. b) Experimental setup: a diode laser beam at λ=1.55μm is incident onto the sample with y-polarization. The light scattered by the metasurface in x-polarization is isolated with a polarizer. A detector mounted on a 3-axis motorized translational stage collects the light passing through a pinhole, attached to the detector, with an aperture of 50μm c) SEM image of the fabricated lens with 3cm focal distance (left). The corresponding phase shift profile calculated from Eq 1 and discretized according to the phase shifts of the 8 antennas is displayed on the right. Inset: close up of patterned antennas. The distance between two neighboring antennas is fixed at Δ=750nm in both directions for all the devices.

To facilitate the design of the metasurfaces, we used a simple analytical model based on the dipolar emitters [27]. The emission of our antennas can be well approximately by that of electric dipoles [28, 29]. We can calculate the intensity of the field ($|E|^2$) scattered from a metasurface for a particular distribution of amplitudes and phases of the antennas by superposing the contributions from many dipolar emitters. This approach offers a convenient alternative to time-consuming FDTD simulations. The metasurface is modeled as a continuum of dipoles with identical scattering amplitude and a phase distribution given by Eqs. 1 and 2. By comparing calculations based on this model and the experimental data, we can determine whether the phase discretization and the slight variations in the scattering amplitudes of the 8 elements create substantial deviations from the operation of ideal devices. The measured far-field for the lens with 3 cm focal distance and the corresponding analytical calculations are presented in Figs. 3 a-c). The results for an ideal axicon and for the axicon metasurface are presented in Figs. 3 d-f. Note that the actual non-diffracting distance of the axicon metasurface is shorter than the ideal DOF because the device is illuminated with a collimated Gaussian beam instead of a plane wave [30].

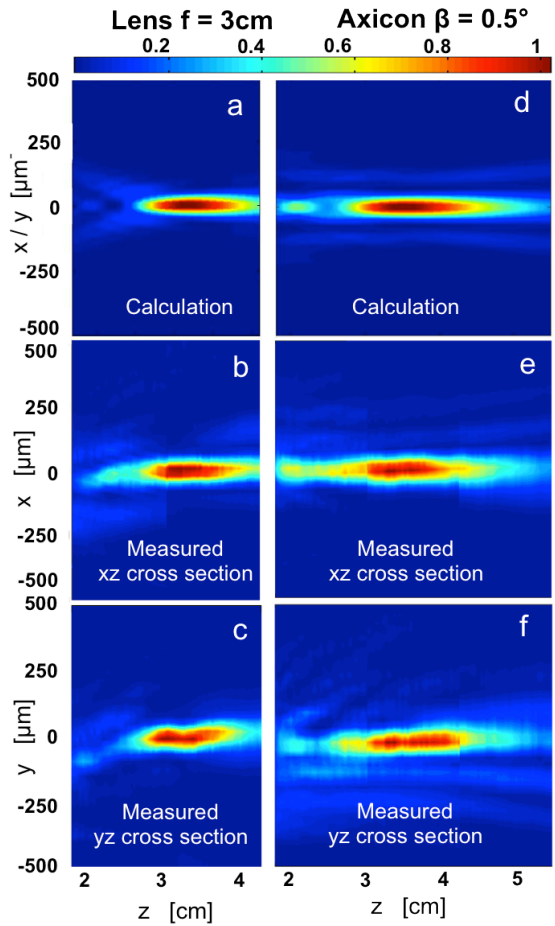

**Fig 3.** a-c) Theoretical calculations and experimental results of the intensity distribution in the focal region for the flat lens with f=3cm. a) is calculated using the dipolar model. b and c are the experimental results showing the xz and yz longitudinal cross section of the 3-dimensional far-field distributions. d-f) Theoretical calculations and experimental results of the intensity distribution for the planar axicon with β=0.5°.

In Fig 4, we present the calculated and the measured intensity profiles in the transverse direction for the three devices. For the lenses, we choose the focal planes to be at z=6 cm (Fig. 4a,d,g) and z=3 cm (Fig. 4b,e,h). For the axicon, the tranverse cross section was taken at a distance of 3.5 cm from the interface which is within the DOF (Fig. 4c,f,i).

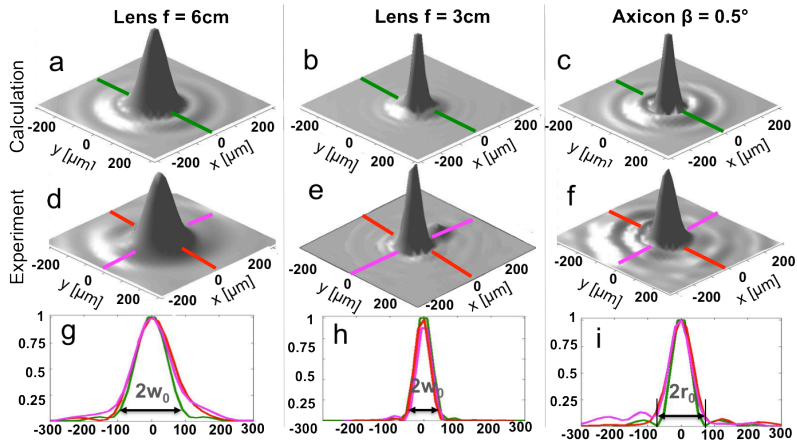

**Fig 4.** a-c) Transverse cross-section of the intensity profiles calculated using the analytical model for the 6cm focal lens (a), the 3cm focal lens (b) and the axicon (c). d-f) are the measured transverse cross-sections of the intensity profiles for the 6cm focal lens (d), 3cm focal lens (e) and axicon (f). In a and b the transverse sections are taken at the focal planes at z=6 cm and z=3 cm respectively. In c it is taken at z=3.5 cm. g-i) The green curves are the line scans of (a)-(c); the red and magenta lines are the line scans of (d)-(f). The beam waists ($w_0$) and the radius of the central lobe ($r_0$) are shown for the focused Gaussian beams and the Bessel beams respectively. The experimental setup used in these measurements is the same as Fig. 2. For measurements in Fig. 4e-h, we used a 15 µm-aperture pinhole. Note the good agreement between calculations and experiments. We estimate that the width of the central lobes for measured transverse sections differ from the expected values by a 5-10% which is attributed to the finite size of the pinhole in front of the detector. The measured profiles display the difference between the Airy patterns generated by the lenses and the Bessel beam created by the axicon.

To prove the possibility of creating lenses with high NA we performed FDTD simulations of the metasurfaces. Instead of the whole lens comprising a 2D array of antennas, we simulated only the unit cell (Fig 5a). This simplified design is equivalent to a cylindrical lens; it is useful for understanding the focusing proprieties of a high numerical aperture objective. Figures 5b and c show the cross-sections of the intensity at the focal plane of the lenses with NA=0.015 (as in the fabricated device) and NA = 0.77.

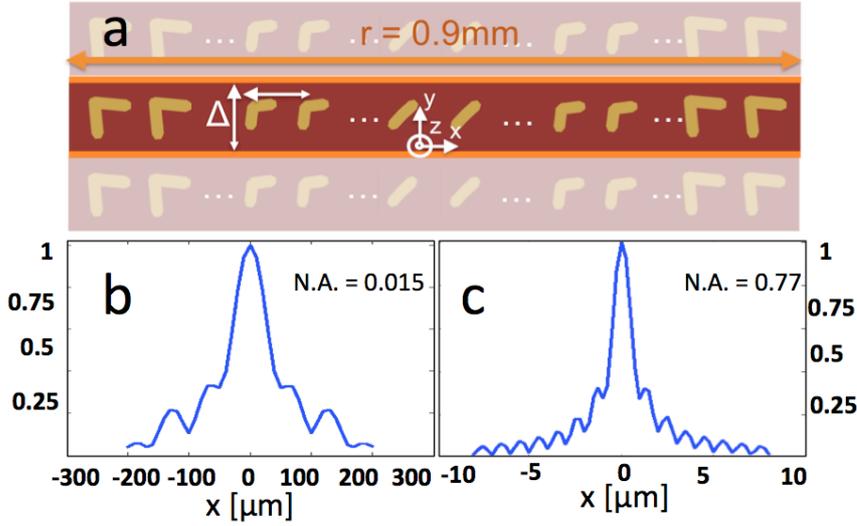

**Fig 5.** a) Schematic of the simulated unit cell. The simulated area is 0.9mm long and the antennas are separated by a distance Δ=300nm. We used Eq.1 to create a distribution of phase shifts for focusing light at the distance f. In the y-direction we use periodic boundary conditions as indicated by the orange lines. In this way the phase modulation is present only in the x-direction and the effect will be the same as that of a cylindrical lens. b-c) Cross-sections at the focal plane for the f=3cm (NA = 0.015) and the f=371µm (NA= 0.77) lenses. The beam waists are $w_0$=50µm and $w_0$=1µm, respectively.

The design of this new class of focusing devices is free from monochromatic-aberrations typically present in conventional refractive optics. The phase distribution created from a spherical lens focuses the light to a single point only in the limit of paraxial approximation; a deviation from this condition introduces monochromatic aberrations such as spherical aberrations, coma and astigmatism [1]. To circumvent these problems, complex optimization techniques such as aspheric shapes or multi-lens designs are implemented [1, 2]. In our case, the hyperboloidal phase distribution imposed at the interface produces a wavefront that remains spherical even for non-paraxial conditions. This will lead to high NA focusing without aberrations.

A present limit of our design is the focusing efficiency, limited to ~10%. This is mainly due to the reflection inside the substrate and the optical losses, which are estimated using FDTD simulations to be approximately 40% and 20%, respectively. The optical losses can be reduced using low-loss metals [31, 32] or other plasmonic materials [33]. The efficiency can be further improved by using impedance matching techniques (e.g. antireflective coatings) and exploiting antenna designs with higher scattering amplitude (e.g. antennas with a metallic back plane operating in reflection mode).

In conclusion, flat lens and axicon designs based on plasmonic metasurfaces are presented. They are characterized by lack of monochromatic aberration even at high NA. We have fabricated and demonstrated two lenses with centimeter scale focal lengths and an axicon with angle $\beta=0.5°$. The experimental results are in good agreement with analytical calculations using a dipolar model. Ultra thin and high NA lens may find applications in microscopy or in other imaging tools. Planar lenses and axicons can be designed for other spectral regions and may become particularly interesting in the mid-infrared, the terahertz and those ranges of frequencies where transparent refractive materials are harder to find compared to the near-infrared and the visible.

Although the present design is diffraction limited, focusing and imaging below the diffraction limit in the far field can be achieved using plates patterned with structures that provide subwavelength spatial resolution of the phase and amplitude of light [8]. Optical phase discontinuities may find applications in such microscopy techniques with super resolution.

[1] Hecht, E. Optics, 3rd ed.; Addison Wesley Publishing Company: Reading, MA, 1997

[2] Kinglsake, R. *Lens Design Fundamentals*, Academis Press, 1978

[3] Lu, F.; Sedgwick F. G.; Karagodsky V.; Chase C.; Chang-Hasnain C. J. C. *Opt. Expr.*, **2010**, 18 12606-12614

[4] McGrath, D. T. *IEEE Trans. on Ant. and Prop*. **1986**, Ap-34, 46-50.

[5] Pozar, D. M. *Electr. Lett.* **1996** 32, 2109-2111.

[6] Huang, F.M; Kao, T.S.;Fedotoc, V. A.; Chen, Y.; Zheludev, N. *Nano Lett.* **2008**, 8, 2469-2472.

[7] Huang, F.M.; Zheludev, N. *Nano Lett.***2009**, 9, 1249-1254.

[8] Rogers, E.T.F.; Lindberg, J.; Roy, T.; Savo, S.; Chad, J.E.; Dennis, M.R.; Zheludev, N.I. *Nat. Mat.*, **2012**, 11, 432-435.

[9] Fattal D.; Li J.; Peng Z.; Fiorentino M.; Beausoleil R. G. *Nat. Phot.* **2010**, 4, 466-470

[10] Verslegers, L,; Catrysse, P.B.; Yu, Z.; White, J.S.; Barnard, E.S.; Brongersma, M.L.; Fan, S. *Nano Lett.* **2009**, 9, 235-238.

[11] Pendry, J. B.; *Phys. Rev. Lett.* **2000** 85, 3966-3969

[12] Smith, D. R.; Pendry, J. B.; Wiltshire, M. C. K. *Science* **2004** 305, 788

[13] Liu, Z.; Lee, H.; Xiong, Y.; Sun, C.; Zhang, X. ***Science***, **2007**, 315, 1686

[14] Cai, W.; Shalaev, V. *Optical Metamaterials Fundamentals and Applications* **2010** Springer

[15] Yu, N.; Genevet, P.; Kats, M. A.; Aieta, F.; Tetienne, J.P.; Capasso, F.; Gaburro, Z. *Science* **2011** 334, 333-337.

[16] Pozar, D. M.; Targonski, S. D.; Syrigos, H. D. *IEEE Trans. Antenn. Propag.* **1997**, 45, 287

[Encinar2001] Encinar, J. A.; *IEEE Trans. Antenn. Propag*. **2001**, 49, 1403

[17] Ryan, C. G. M. et al*., IEEE Trans. Antenn. Propag*. **2010**, 58, 1486

[18] Padilla, P.; Muñoz-Acevedo, A. ; Sierra-Castañer, M.; Sierra-Pérez, M. *IEEE Trans. Antenn. Propag*. **2010**, 58, 2571

[19] Aieta, F.; Genevet, P.; Yu. N.; Kats, M. A.; Gaburro, Z.; Capasso, F. *Nano. Lett.* **2012**, 12, 1702–1706.

[20] Ni, X.; Emani, N. K.; Kildishev, A. V.; Boltasseva, A.; Shalaev, V. M. *Science* **2012**, 335, 427.

[21] Genevet, P.; Yu, N.; Aieta, F.; Lin, J.; Kats, M. A.; Blanchard, R.; Scully, M. O.; Gaburro, Z.; Capasso, F. *Appl. Phys. Lett*. **2012**, 100, 013101.

[22] McLeod, J.H. *J. Opt. Soc. Am*. **1953**, 44, 592-597.

[23] Manek, I.; Ovchinnikov, Yu.B.; Grimm, R. *Opt. Comm*. **1998** 147, 67-70

[24] McLeod, J.H. *J. Opt. Soc. Am*. **1960**, 50 166-169.

[25] Ren. Q.; Birngruber, R. *IEEE J. Quant. Electr*. **1990**, 26, 2305-2308.

[26] Artl, J.; Garces-Chavez, V.; Sibbett, W.; Dholakia, K. *Opt. Comm.* **2001**, 197, 239-245.

[27] Tetienne, J.P.; Blanchard, R.; Yu N.; Genevet, P.; Kats, M.A.; Fan J.A.; Edamura, T.; Furuta, F.; Yamanishi, M.; Capasso, F. *New Journal of Physics* **2011** 13 053057

[28] Kats, M. A.; Genevet, P.; Aoust, G.; Yu. N.; Blanchard, R.; Aieta, F.; Gaburro, Z.; Capasso, F. *PNAS* **2012** XXX, XXX-XXX

[29] Blanchard, R.; Aoust, G.; Genevet, P.; Yu, N.; Kats, M.A.; Gaburro, Z.; Capasso, F. *Phys. Rev. B* **2012,** 85, 155457

[30] Jarutis, V.; Paskauskas, R.; Stabinis, A. *Opt. Comm.* **2000** 184, 105-112

[31] Bobb, D. A. ; Zhu, G. ; Mayy, M.; Gavrilenko, A. V.; Mead, P.; Gavrilenko, V. I.; Noginov, M. A. *Appl. Phys. Lett.* **2009**, 95, 151102

[32] West, P.R.; Ishii, S.; Naik, G.V.; Emani, N.K.; Shalaev, V.M.; A. Boltasseva *Laser & Photon. Rev.* **2010** 4, 795–808

[33] Koppens, F. H.L.; Chang, D. E.; de Abajo, F. J. G. *Nano Lett.* **2011**, 11, 3370–3377

**Acknowledgements**: The authors acknowledge support from the National Science Foundation, Harvard Nanoscale Science and Engineering Center (NSEC) under contract NSF/PHY 06-46094, and the Center for Nanoscale Systems (CNS) at Harvard University. P.G acknowledges funding from the Robert A. Welch Foundation (A-1261). Z. G. acknowledges funding from the European Communities Seventh Framework Programme (FP7/2007-2013) under grant agreement PIOF-GA-2009-235860. M.A.K. is supported by the National Science Foundation through a Graduate Research Fellowship. CNS is a member of the National Nanotechnology Infrastructure Network (NNIN).